\documentclass[usenatbib]{mn2e}
\usepackage{graphicx}
\usepackage{astrobib_mnras2e}
\usepackage{fixltx2e}
\usepackage{times}

\begin{document}
\topmargin-1cm

\newcommand\approxgt{\mbox{$^{>}\hspace{-0.24cm}_{\sim}$}}
\newcommand\approxlt{\mbox{$^{<}\hspace{-0.24cm}_{\sim}$}}
\newcommand{\be}{\begin{equation}}
\newcommand{\ee}{\end{equation}}
\newcommand{\bea}{\begin{eqnarray}}
\newcommand{\eea}{\end{eqnarray}}
\newcommand{\lexp}{\mathop{\langle}}
\newcommand{\rexp}{\mathop{\rangle}}
\newcommand{\rexpc}{\mathop{\rangle_c}}
\newcommand{\pcl}{pseudo-$C_{l}$~}
\newcommand{\pclc}{Pseudo-$C_{l}$~}
\newcommand{\ylm}{Y_{lm}}
\newcommand{\plm}{\psi_{lm}}
\newcommand{\thi}{\hat{\theta}_{i}}
\newcommand{\thj}{\hat{\theta}_{j}}
\newcommand{\nbar}{\bar{n}}
\newcommand{\talm}{\tilde{a}_{lm}}
\newcommand{\tcl}{\tilde{C}_{l}}
\newcommand{\nhat}{\hat{\bf n}}
\newcommand{\kvec}{{\bf k}}
\newcommand{\khat}{\hat{\bf k}}
\newcommand{\zmax}{Z_{\rm max}}
\def\bi#1{\hbox{\boldmath{$#1$}}}

\title{A Robust Estimator of the Small-Scale Galaxy Correlation Function}
\author[Padmanabhan et al.]
{Nikhil Padmanabhan$^{1,4}$\thanks{NPadmanabhan@lbl.gov},
Martin White$^{2}$,
Daniel J. Eisenstein$^{3}$\\
$^{1}$ Physics Division, Lawrence Berkeley National Laboratory,
1 Cyclotron Rd., Berkeley, CA 94720, USA. \\
$^{2}$ Department of Physics and Astronomy, 601 Campbell Hall,
University of California Berkeley, CA 94720, USA.\\
$^{3}$ Steward Observatory, University of Arizona, 933 N. Cherry Ave,
Tucson, AZ 85721, USA. \\
$^{4}$ Hubble Fellow, Chamberlain Fellow \\
}
\date{\today}
\maketitle

\begin{abstract}
We present a new estimator, $\omega$, of the small scale galaxy
correlation function that is robust against the effects of redshift
space distortions and large scale structures. The estimator is a
weighted integral of the redshift space or angular correlation function
and is a convolution of the real space correlation function with a
localized filter. This allows a direct comparison with theory,
without modeling redshift space distortions and the large
scale correlation function. This
has a number of advantages over the more traditional $w_{p}$
estimator, including  
(i) an insensitivity to large scale structures
and the details of the truncation of the line of sight integral,
(ii) a well localized kernel in
$\xi(r)$, and (iii) being unbinned. We discuss how this estimator would be used in practice,
applying it to a sample of mock galaxies selected from the Millennium
simulation. 
\end{abstract}



\section{Introduction}
\label{sec:introduction}

The two point correlation function of galaxies is 
one of the foremost probes of the physics of galaxy formation and 
evolution. Unfortunately, peculiar velocities smear out the galaxy
distribution along the line of sight, causing the observed
3D correlation function, $\xi_{s}(r)$, 
to deviate from the underlying isotropic correlation function, $\xi(r)$.
The most commonly employed solution has been to consider the projected
correlation function \citep{1983ApJ...267..465D}, 
\begin{equation}
\label{eq:wpdef}
w_p(R) \equiv \int_{-\infty}^{\infty} dZ\ \xi_{s}\left(R,Z\right)
\end{equation}
where $R$ and $Z$ are the transverse and line of sight coordinates respectively.
 
Although this estimator has enjoyed widespread use 
\citep[eg. see][for recent applications to the 2dFGRS,SDSS and DEEP2 redshift surveys]
{2003MNRAS.346...78H,2005ApJ...621...22Z,2006ApJ...644..671C}, 
it has an important drawback. 
While the integral formally extends over the entire 
$Z$ axis, it is truncated at $\zmax$, where $\zmax \gg R$ is typically 
several tens of Mpc to avoid biases. This mixes in a large
range of scales, strongly correlating measurements and making the 
estimator sensitive to possibly poorly measured long wavelength modes.
Furthermore, the quantity of interest from the perspective
of galaxy formation and evolution is $\xi(r)$ where $r$ is determined
by the size of dark matter halos ($\sim 1~{\rm Mpc}$). Estimating this 
from $w_{p}$ requires disentangling it from 
large scale correlations. 

All of these problems can be solved by inverting the Abel integral
to obtain $\xi(r)$,
\begin{equation}
\label{eq:wp2xi}
\xi(r) = - \frac{1}{\pi} \int_{r}^{\infty} \frac{dR}{\sqrt{R^{2}-r^{2}}}
\frac{d w_{p}}{dR} \,\,.
\end{equation}
Unfortunately, the resulting $\xi(r)$ has severe anti-correlations between
bins, characteristic of all deconvolution problems. Furthermore, these get 
worse as one reduces the bin sizes to avoid binning effects. To avoid these 
problems, it has become popular to attempt to theoretically model $w_{p}$;
however, one is then left with the disadvantages of $w_{p}$.

Motivated by this, we suggest a new statistic $\omega$ that is a more
robust estimate of the small and intermediate 
scale ($\sim 1-10~{\rm Mpc}$) correlation 
function. As the problematic modes in $w_{p}$ are slowly varying, we 
suggest high-pass filtering to remove them (Sec.~\ref{sub:theory}). 
Sec.~\ref{sub:family} then describes a family of possible filters
chosen to add the
additional desirable properties of being compact in the projected correlation function and 
well localized in the
3D correlation function. Sec.~\ref{sub:computational} describes how to estimate $\omega$ and
demonstrates that it can be computed without any need to (arbitrarily) bin the data.
Sec.~\ref{sub:testing} then tests these ideas with a sample of galaxies 
drawn from the Millennium simulation. We summarize in Sec.~\ref{sec:discussion}.

\section{A New Clustering Statistic : $\omega$}
\label{sec:omega}

\subsection{Theory}
\label{sub:theory}

If we truncate
the integral in Eq.~\ref{eq:wpdef} at $\pm \zmax$, we introduce an error in our
$w_{p}$ estimator of
\begin{equation}
\label{eq:errorwp}
\Delta w_{p}(R) = 2 \int_{\zmax}^{\infty} dZ\, \xi_{s} \left( R,Z \right)
\qquad .
\end{equation}
It is important to emphasize that the correlation function that appears 
above is the redshift-space correlation function, and predicting it 
requires a complete model of redshift-space distortions, not just 
the isotropic correlation function.
However, since for $\zmax \gg R$, this error is a slowly varying function of $R$, 
one can reduce it by high-pass filtering,
\begin{eqnarray}
\label{eq:omegadef}
\omega(R_s) & \equiv & 2\pi \int R\,dR\ G(R, R_{s}) w_{p}(R)  \nonumber \\
& = & 2\pi \int R\,dR\ G(R, R_{s})
\int^{\zmax}_{-\zmax} \,dZ\ \xi_{s}(R,Z) \,\,
\end{eqnarray}
where the filter $G(R,R_{s})$ is designed to be compact, compensated
($\int R\,dR\ G(R)=0$) and have a characteristic scale $R_s$.  
A compensated filter reduces the influence of slowly varying
modes, coming from large $r$ in $\xi(r)$, while compactness avoids
extrapolations of the data. We further assume $G(R)$ has units
of inverse volume, making $\omega$ dimensionless.

Using the above definitions, we can relate $\omega(R_{s})$ to the 
real space correlation function $\xi(r)$,
\begin{equation}
  \label{eq:omegadef_xi}
  \omega(R_s) = 4\pi \int dr\,r^{2}  \xi(r)W(r)
\end{equation}
where
\begin{equation}
  \label{eq:wfilter_def}    
  W(r) = \frac{1}{r}\int_0^r \frac{R\,dR}{\sqrt{r^2-R^2}}\ G(R)
  \qquad .
\end{equation}
Note that Eqs.~\ref{eq:omegadef}-\ref{eq:wfilter_def} offer a natural
generalization of Eq.~\ref{eq:wp2xi}, with a particular ``binning''
that smooths over the anti-correlations in $\xi$ which come from inversion.

In principle, one could generalize the above by considering 
a filter in $R$ and $Z$, i.e.
\begin{equation}
  \omega = 2\pi \int RdR\,dZ\,G(R,Z,R_{s}) \xi_{s}(R,Z)\,\,,
\label{eq:general_omega}
\end{equation}
where $G(R,Z,R_{s})$ is chosen to maximize the $S/N$ under 
the assumption of smoothness of the redshift distortions in $Z$.
A simple example would be to parameterize the $k_z$ dependence
of the power spectrum by a low-order polynomial and extrapolate
the fit to $k_z=0$.  After much experimentation with basis choices,
apodization and weighting we were unable to find an acceptable filter
that converged to a real space statistic better than trivially weighting
in $Z$, so we stick with our original unweighted integral.
Note however that one trivial generalization would be to allow $\zmax$
to vary with $R_{s}$.  Whether or not this is useful will depend on
details of the galaxy sample, and so we simply mention the possibility
and defer experimentation to actual applications to real data.

Finally, we observe that a similar formalism applies to the angular
correlation function.  If we consider a galaxy sample with redshift
distribution $\phi$, the angular correlation function $w(\theta)$ is
related to the 3D correlation function by 
Limber's equation \citep{1953ApJ...117..134L,1999coph.book.....P}
\footnote{Assuming
that we are probing scales much less than the width of the redshift
distribution.},
\begin{equation}
\label{eq:limber}
  w(\theta) =  \int_{0}^{\infty} d\chi\ \phi^2(\chi) \int dZ\,
  \xi\left(\sqrt{\left(\theta \chi\right)^{2}+Z^{2}}\right) \,\,,
\end{equation}
where $\chi$ is the radial distance and $\int \phi\, d\chi=1$.
We now proceed as above defining
\begin{equation}
\label{eq:gtheta}
  \omega(\theta_{s}) = 2\pi\int \theta d\theta\,G(\theta, \theta_{s})
  w(\theta)\,\,,
\end{equation}
which probes scales around $\theta_{s} \chi_{0}$, with $\chi_0$ the
characteristic distance of the galaxy sample.
As before, this is related to the 3D correlation function by
Eq.~\ref{eq:omegadef_xi} with $W(r)$ replaced by $W_{\theta}(r)$,
\begin{equation}
\label{eq:Wtheta}
  W_{\theta}(r) = \int_{0}^{\infty} d\chi
    \left(\frac{\phi}{\chi}\right)^{2} W\left(\frac{r}{\chi}\right) \,\,,
\end{equation}
where $W$ is given by Eq.~\ref{eq:wfilter_def} with the integration variable
interpreted as an angular coordinate.
Note that this is the same window function as for $w_{p}$ except that it is
now smoothed by $\phi$.

\subsection{A Family of Filters}
\label{sub:family}

\begin{figure}
\begin{center}
\leavevmode
\includegraphics[width=3.0in]{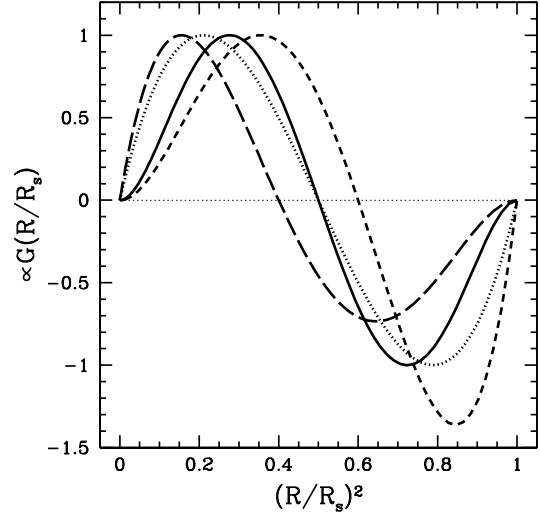}
\end{center}
\caption{Examples of $G(R/R_{s})$ chosen from the two parameter family
described in the text, $(\alpha, \beta)$=(1,1) [dotted], (2,1) [short-dashed],
(1,2) [long-dashed], and (2,2) [solid]. We adopt the (2,2) filter throughout
this paper. The $x-$axis is chosen such that the area under the curves is
zero, while the filter normalization is arbitrary. Note that $\alpha$ controls
the small-$R$ behaviour of the filter, while $\beta$ determines its derivative
at $R \sim R_{s}$.}
\label{fig:gfilter}
\end{figure}

\begin{figure}
\begin{center}
\leavevmode
\includegraphics[width=3.0in]{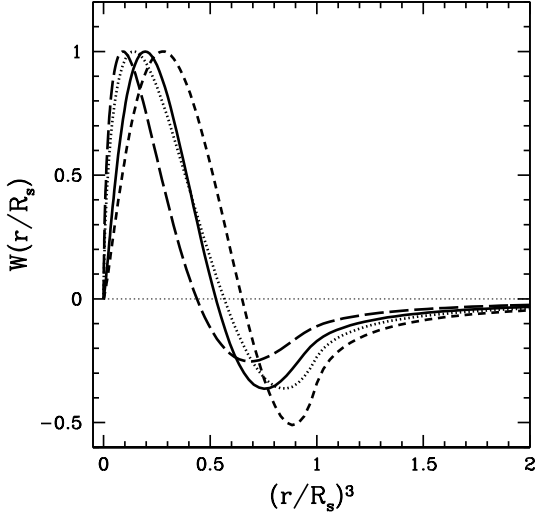}
\end{center}
\caption{Corresponding $W(r/R_{s})$ curves for the filters shown in
Fig.~\protect\ref{fig:gfilter}. The $x-$axis is chosen to correspond to
the appropriate measure in the integral in Eq.~\ref{eq:omegadef_xi}. 
Note that the large $r$ behaviour
for all these filters is $r^{-4}$, corresponding to a singly compensated
$G$ filter.}
\label{fig:Wfilter}
\end{figure}

A simple two parameter family of filters that satisfy the requirements
to be compact and compensated and that
are analytically tractable are $G(R)=R_s^{-3} G(x=R/R_s)$ with\footnote{For
$w(\theta)$ the prefactor is $\theta_s^{-3}$ and $x=\theta/\theta_s$.}
\begin{eqnarray}
G(x) = &  x^{2\alpha} (1-x^{2})^{\beta} (c - x^{2}) & x \le 1 \\
= & 0 & x > 1 \,\,,
\label{eq:gfilter}
\end{eqnarray}
where $\alpha, \beta=1,2,...$ and
\begin{equation}
c = \frac{\alpha + 1}{\alpha + \beta + 2} \,\,,
\label{eq:cdef}
\end{equation}
is determined by the requirement that the filter be compensated.
The corresponding real space filters have the form
\begin{eqnarray}
W(y) = & P(y) & y \le 1\\
= & P(y) + \frac{\sqrt{y^{2}-1}}{y} Q(y) & y > 1 \,\,,
\label{eq:wfilter}
\end{eqnarray}
where $P$ and $Q$ are simple polynomials and $y=r/R_s$.
Table~\ref{tab:Wpoly} lists the values for $c$, $P(y)$ and $Q(y)$ for the
first four filters in this family, and
Figs.~\ref{fig:gfilter} and ~\ref{fig:Wfilter}
plot $G(x)$ and $W(y)$ for these filters.

\begin{table*}
\begin{tabular}{ccll}
\hline
$(\alpha, \beta)$ & $c$  & $P(y)$ & $Q(y)$ \\
\hline
(1,1) & $\frac{1}{2}$ &
    $\frac{1}{105}\left[35y^{2} - 84y^{4} + 48y^{6}\right]$ & 
    $\frac{-1}{105}\left[1 + 11y^{2} - 60y^{4} + 48y^{6} \right]$\\
(1,2) & $\frac{2}{5}$ & 
    $\frac{4}{1575}\left[105y^{2} - 378y^{4} + 432y^{6} - 160y^{8}\right]$ & 
    $\frac{-8}{1575} \left[ 1 + 14 y^{2} - 111 y^{4} + 176 y^{6} - 80 y^{8}\right] $ \\
(2,1) & $\frac{3}{5}$ & 
    $\frac{4}{1575}\left[126 y^{4} - 288 y^{6} + 160 y^{8}\right]$ &
    $\frac{-4}{1575} \left[1 + 5y^{2} + 42y^{4} - 208y^{6} + 160y^{8} \right]$ \\
(2,2) & $\frac{1}{2}$ & 
    $\frac{4}{3465}\left[231y^{4} - 792y^{6} + 880y^{8} - 320y^{10}\right]$ &
    $\frac{-4}{3465} \left[ 1 + 6y^{2} + 65 y^{4} - 472 y^{6} + 720y^{8} - 320y^{10} \right]$\\
\hline
\end{tabular}
\caption{\label{tab:Wpoly}  Table of $c$, $P(y)$ and $Q(y)$ for the first
four filters in  the simple two parameter family discussed in
Sec.~\protect\ref{sub:family}.}
\end{table*}

The choice of $\alpha$ and $\beta$ determine the $x \rightarrow 0$ and
$x \rightarrow 1$ behaviour of $G(x)$.
For small $x$, $G(x) \sim x^{2\alpha}$; increasing $\alpha$
reduces the sensitivity of $\omega$  to measurements of the correlation
function on scales much smaller than the scales of interest.
At the other end, $\beta-1$ determines the number of derivatives of $G(x)$
that vanish at $x=1$. Increasing $\beta$ increases the smoothness at $x=1$
and reduces susceptibility to ringing;  however, large $\alpha$
and $\beta$ makes $G(x)$ sharply peaked, increasing its susceptibility to
noise in the correlation function.  After some experimentation, we 
adopted $(\alpha,\beta)=(2,2)$ in what follows as a good compromise.
Note that Eq.~\ref{eq:gfilter} can
be generalized to a $N-$compensated filter 
\begin{equation}
\label{eq:gfilter_mult}
G_{N}(x) = x^{2\alpha} (1-x^{2})^{\beta} \prod^{N}_{i} (c_{i} - x^{2}) \,\,,
\end{equation}
where the $c_{i}$ are determined by the compensation conditions
\begin{equation}
  \int R\,dR\ R^{2j}G(R) = 0 \quad {\rm for}\ 0\le j<N\quad .
\end{equation}
For an $N$-compensated filter $W(r)\sim r^{-2(N+1)}$ as $r\to\infty$.
However, since the galaxy correlation function is  falling off steeply with
increasing radius, we find that a singly compensated filter is sufficient
to remove the large radius contribution to $\omega$.  For example, if
$\xi\sim r^{-2}$, $\omega(R_s)$ converges to $1\%$ by $r\simeq 2 R_s$.
Furthermore, as an $N$-compensated filter has $N$ nodes, a multiply compensated
filter is rapidly oscillating and therefore sensitive to noise. For
these reasons, we recommend using a singly compensated filter.
Parenthetically, we note that the polynomial representations of $W(y)$ in
Table~\ref{tab:Wpoly} are not numerically ideal for $y \gg 1$. We therefore 
suggest switching to a series expansion in $1/y$ for large $y$; for convenience, 
we give the expansion for the (2,2) filter here,
\begin{equation}
\label{eq:w22series}
W(y) = -\left[\frac{1}{3360y^{4}} + \frac{1}{4480y^{6}} + \frac{5}{32256y^{8}} 
+ {\cal O}\left(\frac{1}{y^{10}}\right)\right]\,\,.
\end{equation}
As expected for a singly compensated filter, the leading term is of order $y^{-4}$.

The above discussion has focussed on a real space description; it is 
however instructive to relate $\omega$ to the 3D real space power spectrum, $P(k)$.
We recall that $w_{p}(R)$ is a Hankel transform of the power spectrum,
\begin{equation}
\label{eq:hankel}
w_{p}(R) = \pi R \int \frac{dk}{k}\, \Delta^{2}(k) \frac{J_{0}(kR)}{kR}\,\,,
\end{equation}
where we have used the isotropy of the real-space power spectrum and
$\Delta^{2}(k) = k^{3} P(k)/2\pi^2$. Using 
Eq.~\ref{eq:omegadef}, we obtain 
\begin{equation}
\label{eq:omegak}
\omega = \int \frac{dk}{k}\, \Delta^{2}(k) \widetilde{W}(k) \quad ,
\end{equation}
where 
\begin{equation}
\label{eq:wtilde}
\widetilde{W}(k) = \frac{2\pi^2}{k}\int_{0}^{R_{s}} RdR\, G(R)J_{0}(kR) \,\,.
\end{equation}
Although this integral can be done analytically for any polynomial $G(R)$, 
we simply note that for $kR_{s} \ll 1$, $\widetilde{W}(k) \sim k^{2N-1}$, 
yet another manifestation of the insensitivity of $\omega$ to large scales 
(small $k$). In particular, $\widetilde{W}(0)=0$, implying that $\omega$ 
is insensitive to the mean density of the sample and therefore, the integral
constraint.

\begin{figure}
\begin{center}
\leavevmode
\includegraphics[width=3.0in]{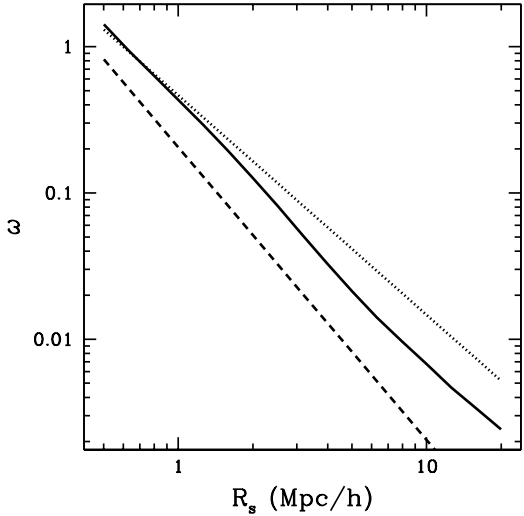}
\end{center}
\caption{Some examples of $\omega$ (assuming a (2,2) filter) :
for $\xi(r) = (r/10 [{\rm Mpc}/h])^{-1.5}$ [dotted], 
for $\xi(r) = (r/3 [{\rm Mpc}/h])^{-2}$ [dashed],
and for a sample of galaxies (see text) from the Millennium simulation [solid].
}
\label{fig:omega}
\end{figure}

We conclude this section with the exact analytic expression for $\omega$,
assuming a power law correlation function,
\begin{equation}
\label{eq:powerxi}
\xi(r) = \left( \frac{r}{r_{0}}\right)^{-\gamma} 
\qquad {\rm for}\ \gamma>1\ .
\end{equation}
Integrating along the line of sight, one obtains
\begin{equation}
\label{eq:powerwp}
w_{p}(R) = \frac{\sqrt{\pi}
  \Gamma(\frac{\gamma-1}{2})}{\Gamma(\frac{\gamma}{2})} r_{0} 
  \left( \frac{R}{r_{0}}\right)^{-\gamma+1} \,\,.
\end{equation}
Integrating along $R$, one obtains (for the (2,2) filter),
\begin{equation}
\label{eq:poweromega}
\omega = \frac{2 \pi^{3/2} \Gamma(\frac{\gamma-1}{2})}{\Gamma(\frac{\gamma}{2})}
\frac{4(\gamma-1)}{(7-\gamma)(9-\gamma)(11-\gamma)(13-\gamma)}
 \left( \frac{R}{r_{0}}\right)^{-\gamma} \,.
\end{equation}
Note that $\omega(R_{s}) \propto \xi(R_{s})$, allowing one to trivially
translate between $\xi$ and $\omega$ for the case of power law correlation
functions.  For $\gamma\simeq 2$ the prefactor is a few per cent, so $\omega$
will be order unity on Mpc scales for modestly biased populations at $z=0$
(Fig.~\ref{fig:omega}).

\subsection{Computational Considerations}
\label{sub:computational}

The correlation function $\xi_{s}(R,Z)$ can be estimated via
\begin{equation}
\label{eq:xi_estimator}
\xi_{s}(R,Z) = \frac{DD(R,Z)}{RR(R,Z)} -1  \,\,,
\end{equation}
where $DD(R,Z)$ is the number of galaxy pairs separated by $R$ and $Z$, while 
$RR(R,Z)$ is the analogous quantity for randomly distributed points. 
We assume that $DD$ and $RR$ are actual number counts in a bin of a finite
size, as opposed to density distributions. 
Furthermore, although the number of random points is much greater than the 
number of data points $n_{D}$, we assume that $RR$ is normalized to the 
number of data pairs, $n_{D}^{2}$.
Our new statistic
is simply an integral of the correlation function,
\begin{equation}
\label{eq:omega_estimate}
\omega(R_{s}) = 2 \pi \int_{0}^{R_{s}} dR \int_{-\zmax}^{\zmax} dZ\,R G(R,R_{s}) 
    \xi_{s}(R,Z) \,\,.
\end{equation} 
Substituting Eq.~\ref{eq:xi_estimator} into the above
(compare with Eq.~\ref{eq:omegadef}), we obtain
\begin{equation}
\label{eq:omega_estimate2}
\omega(R_{s}) = 2 \pi \int dR\, R G(R,R_{s}) \int_{-\zmax}^{\zmax} dZ \, \frac{DD(R,Z)}{RR(R,Z)} \,\,,
\end{equation}
where the constant piece vanishes due to the compensated nature of $G(R)$,
a manifestation of our insensitivity to the integral constraint. 

When estimating small-scale correlations, it is common to use small
bins in $R$, which requires large random samples to compute $RR$ accurately
in each individual bin.  This is particularly true because $\xi \gg 1$ 
means that one needs random sets far larger than the data set to keep
the shot noise in $RR$ smaller than that in $DD$.  Such $RR$ summations 
easily dominate the computational time. This would be a serious problem for 
a naive implementation of Eq.~\ref{eq:omega_estimate2}, since the integral 
requires narrow bins in $Z$ and $R$ to avoid binning related errors.
However, we can make an important simplification by noting that $RR$ is a purely
geometrical quantity that depends only on the survey geometry and the number of 
data points. 
We can therefore 
write it as \footnote{assuming certain mathematical niceties 
(eg. the survey geometry is not fractal) that are always true
for real observations},
\begin{equation}
\label{eq:RRdef}
RR(R,Z) = 2 \pi R^{2} \bar{n}_{D} n_{D} \Phi(R,Z) \Delta \ln R \Delta Z \,\,.
\end{equation}
where $\bar{n}_{D}$ is the density of data points and the survey geometry is
encoded in $\Phi(R,Z)$. The advantage of this reformulation  is that the form
of $\Phi(R,Z)$ is highly constrained : $0 \le \Phi(R,Z) \le 1$,
$\Phi(R,Z) \rightarrow 1$ as $R,Z \rightarrow 0$ and it is both smooth and
slowly varying for scales smaller that the survey size.
This allows us to measure $\Phi(R,Z)$ by a simple Monte Carlo integration
analogous to $RR$, but with significantly smaller random samples (for a
given shot noise error), since we can smooth over many bins.
For surveys where the angular and radial selection functions are separable,
we can gain even further by separating $\Phi(R,Z) = \phi_R(R)\phi_Z(Z)$,
allowing us to project along each dimension and thus increase our statistics.

The above allows us to consider arbitrarily fine bins in $R$ and $Z$ without
the usual shot noise penalty in computing $RR$.  We therefore imagine a
binning such that there is either zero or one $DD$ pair per bin.
Substituting this into Eq.~\ref{eq:omega_estimate2}, we can approximate the
integral by a Riemann sum; this transforms the integral into a sum over
pairs in $DD$,
\begin{equation}
\label{eq:omega_riemann}
\omega(R_{s}) = \sum_{i \in DD} 
\frac{G(R_{i})}{\bar{n}_{D} n_{D} \Phi(R_{i}, Z_{i})} H(R_{i},Z_{i}) \,\,,
\end{equation} 
where
\begin{equation}
\label{eq:hdef}
H(R_{i},Z_{i}) \equiv \Theta(R_{s}-R_{i}) \Theta(\zmax - |Z_{i}|)
\end{equation}
such that the Heaviside functions restrict the sum to pairs with $R < R_{s}$
and $|Z|<\zmax$.  This avoids the need to ever bin the data\footnote{We note
as an aside that in a periodic simulation cube $\Phi\equiv 1$ and $\omega$
reduces to a sum over pairs weighted by $G(R_i)$.}.
Finally, we note that the above is trivially generalized to the Landy-Szalay
\citep{1993ApJ...412...64L} correlation function estimator,
\begin{eqnarray}
\lefteqn{\omega_{LS} = \omega_{DD/RR}} \nonumber \\
\label{eq:omega_riemann_ls}
&&\mbox{}  - 2 \sum_{j \in {\rm DR}} 
\frac{G(R_{j})}{\bar{n}_{D} n_{D}\Phi(R_{j}, Z_{j})} H(R_{i},Z_{i}) \,\,,
\end{eqnarray}
where $\omega_{DD/RR}$ is given by Eq.~\ref{eq:omega_riemann}, and $j$ runs
over all data-random pairs.  Note however that this second summation is
computationally expensive for the reasons mentioned above. 

\subsection{Testing the Filter}
\label{sub:testing}

In order to test the above ideas, we select a sample of galaxies \citep{2006MNRAS.365...11C} from the Millennium simulation
\citep{2005Natur.435..629S} with SDSS $r$ band absolute magnitudes $< -20.0$ at $z=0$, 
corresponding to approximately $M_{*}+0.5$ galaxies with a space density of $\sim 7.2 \times 10^{-3} 
(h/{\rm Mpc})^{3}$. This particular sample of galaxies
was chosen to minimize Poisson noise while approximating a typical sample of galaxies,
but is otherwise arbitrary.
Assuming the distant observer approximation, we translate the galaxies 
to redshift space, and subdivide the simulation volume into fifty $100 \times 100 \times 250 {\rm Mpc}/h$
sub-volumes where the long axis corresponds to the redshift space direction. This yields 150 
(using each of the coordinate axes as the redshift space direction in turn) samples of galaxies
from which we measure $\omega$, as discussed above.

\begin{figure}
\begin{center}
\leavevmode
\includegraphics[width=3.0in]{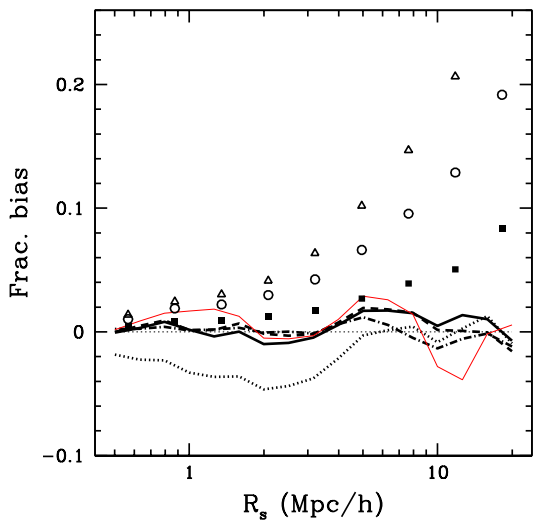}
\end{center}
\caption{The fractional bias in $\omega$ 
(${\rm bias} = (\omega - \omega_{\rm ref})/\omega_{\rm ref}$) obtained from
the redshift space distribution of galaxies in the Millennium simulation
for different $\zmax$ (in Mpc/$h$) --  20 (dotted), 40 (thick solid), 
60 (dashed), and 80 (dot-dashed). The true value ($\omega_{ref}$) is assumed to be 
obtained by integrating to a $\zmax = 100 {\rm Mpc}/h$; the figure shows that
$\omega$ has converged to better than $2\%$ by $\zmax=40 {\rm Mpc}/h$. The 
thin solid (red) line plots $\omega$ obtained by integrating over the real space
correlation function. The fluctuations in $\Delta \omega$ are consistent with 
measurement noise. For comparison, the points show the analogous bias
terms for $w_{p}$, with triangles, circles, and squares corresponding to 
$\zmax$ = $40$, $60$, and $80 {\rm Mpc}/h$ respectively. 
For a fair comparison with $\omega$, we plot the bias in $w_{p}$ at $R_{s}/2$,
which roughly corresponds to the central scale probed by $\omega$.
Note that the biases in
$w_{p}$ are significantly larger than those for $\omega$.
}
\label{fig:bias}
\end{figure}

Fig.~\ref{fig:bias} shows the bias in $w_{p}$ and $\omega$ 
(defined as ${\rm bias}= (f-f_{\rm ref})/f_{\rm ref}$ where $f=w_{p},\omega$)
for the above sample determined
using the estimator in Sec.~\ref{sub:computational}, as a function of transverse scale and $\zmax$.
The reference value is determined by projecting the redshift space correlation function 
to $\zmax = 100 {\rm Mpc}/h$; using this as the reference
mitigates the effects of sample variance and periodicity. The bias in $\omega$
is $< 2\%$ for $\zmax \ge$ 40 Mpc/$h$; smaller scales are still affected by the nonlinear 
redshift space distortions. Furthermore, we observe that our estimates of $\omega$ from the 
redshift space correlation function agree with a direct integration of the isotropic, real
space correlation function.

In contrast with the above, the estimates of $w_{p}$ are both biased at the 5-10 \% level
and show a significantly slower convergence to the true value. For instance, we find it
necessary to integrate to $\zmax=80 {\rm Mpc}/h$ to obtain biases $< 5\%$ 
(still significantly larger than the biases in $\omega$) for $R <  10 {\rm Mpc}/h$. 
This serves to highlight the difficulties in measuring $w_{p}$ and the utility of $\omega$
as an alternate measure of galaxy clustering. Although the 
precise values of the bias will depend on the exact details of the galaxy sample, 
the qualitative aspects of the above is generically true.

\section{Discussion}
\label{sec:discussion}

We present an alternative, $\omega(R_{s})$, to the commonly used $w_{p}(R)$
projected correlation function as a robust measure of the small scale
($\sim {\rm Mpc}$) galaxy correlation function.
This is simply a filtered version of $w_{p}(R)$, and can be straightforwardly
determined by a weighted sum of pairs in the data i.e.~there is no reason to
go through an intermediate step of estimating $w_{p}(R)$.
The features of $\omega$ are :
\begin{enumerate}
    \item {\it Improved Convergence (with $\zmax$) to the real
    space clustering statistic} : It is tempting to believe that one can model the error in $w_{p}$
    introduced by the $\zmax$ truncation, simply by specifying $\xi(r)$ out to large scales (most often
    with the linear theory prediction). We reiterate that this is not true - the error in $w_{p}$ is determined
    by the redshift-space correlation function (Eq.~\ref{eq:errorwp}), and requires a model of redshift-space 
    distortions on all scales. However, recall that it was our uncertainty in redshift-space distortions that led
    us to $w_{p}$ in the first place. 
    
    In contrast, $\omega$ converges to the real-space clustering statistic significantly faster than 
    $w_{p}$ for similar transverse scales, making it insensitive to the precise value of $\zmax$ used.
    Seen in this context, $\omega$ completes the partial removal of redshift space information in $w_{p}$,
    as originally intended.
    
    Furthermore, since $\omega$ converges significantly faster with $\zmax$, it 
    is possible to truncate the underlying $w_{p}$ integral at a lower $\zmax$ than would have been 
    naively possible, eliminating large scale noise and possibly reducing the errors in any downstream 
    quantities derived from the data. The exact details of this are dependent on the exact details of 
    the galaxy sample; we limit ourselves here to pointing out this possibility.
    
    \item {\it Well localized in real space} : The real space filter, $W(r)$,
    is well localized in real space implying that $\omega(R_{s})$ probes a
    relatively narrow range of scales around $R_{s}/2$. 

    \item {\it Immune to the integral constraint} : A corollary to the above is that $\omega(R_{s})$ 
    is immune to errors in the mean density, and therefore is insensitive to the integral constraint.
        
    \item {\it Insensitivity to small scales} : An appropriate choice of $G(R)$ makes $\omega$ insensitive
    to measurements of clustering on very small scales. This is important as it is these scales that are the 
    most sensitive to systematics in galaxy selection.
    
    \item {\it Unbinned} : $\omega(R_{s})$ is a naturally unbinned quantity, removing any need for an arbitrary
    choice of binning.    
\end{enumerate}
Finally, we point out that there is a natural generalization of $\omega$ for angular correlations.

We thank Darren Croton and the Millennium Simulation group for providing us with the catalogue
of galaxies used here. 
Galaxy and dark halo catalogues are publicly available
for the Millennium Simulation at \texttt{http://www.mpa-garching.mpg.de/millennium}.
NP is supported by a NASA Hubble Fellowship HST-HF-01200. MW is supported
by NASA. DJE is supported by NSF AST-0407200. This document is LBNL Report LBNL-62026.
 
\bibliography{biblio,preprints}
\bibliographystyle{mnras}

\end{document}